# Multi-Center Fetal Brain Tissue Annotation (FeTA) Challenge 2022 Results

Kelly Payette, Céline Steger, Roxane Licandro, Priscille de Dumast, Hongwei Bran Li, Matthew Barkovich, Liu Li, Maik Dannecker, Chen Chen, Cheng Ouyang, Niccolò McConnell, Alina Miron, Yongmin Li, Alena Uus, Irina Grigorescu, Paula Ramirez Gilliland, Md Mahfuzur Rahman Siddiquee, Daguang Xu, Andriy Myronenko, Haoyu Wang, Ziyan Huang, Jin Ye, Mireia Alenyà, Valentin Comte, Oscar Camara, Jean-Baptiste Masson, Astrid Nilsson, Charlotte Godard, Moona Mazher, Abdul Qayyum, Yibo Gao, Hangqi Zhou, Shangqi Gao, Jia Fu, Guiming Dong, Guotai Wang, ZunHyan Rieu, HyeonSik Yang, Minwoo Lee, Szymon Płotka, Michal K. Grzeszczyk, Arkadiusz Sitek, Luisa Vargas Daza, Santiago Usma, Pablo Arbelaez, Wenying Lu, Wenhao Zhang, Jing Liang, Romain Valabregue, Anand A. Joshi, Krishna N. Nayak, Richard M. Leahy, Luca Wilhelmi, Aline Dändliker, Hui Ji, Antonio G. Gennari, Anton Jakovčić, Melita Klaić, Ana Adžić, Pavel Marković, Gracia Grabarić, Gregor Kasprian, Gregor Dovjak, Milan Rados, Lana Vasung, Meritxell Bach Cuadra* (IEEE Member), Andras Jakab*

*Abstract*—Segmentation is a critical step in analyzing the developing human fetal brain. There have been vast improvements in automatic segmentation methods in the past several years, and the Fetal Brain Tissue Annotation (FeTA) Challenge 2021 helped to establish an excellent standard of fetal brain segmentation. However, FeTA 2021 was a single center study, and the generalizability of algorithms across different imaging centers remains unsolved, limiting real-world clinical applicability. The multi-center FeTA Challenge 2022 focuses on advancing the generalizability of fetal brain segmentation algorithms for magnetic resonance imaging (MRI). In FeTA 2022, the training dataset contained images and corresponding manually annotated multi-class labels from two imaging centers, and the testing data contained images from these two imaging centers as well as two additional unseen centers. The data from different centers varied in many aspects, including scanners used, imaging parameters, and fetal brain super-resolution algorithms applied. 16 teams participated in the challenge, and 17 algorithms were evaluated. Here, a detailed overview and analysis of the challenge results are provided, focusing on the generalizability of the submissions. Both in- and out of domain, the white matter and ventricles were segmented with the highest accuracy, while the most challenging structure remains the cerebral cortex due to anatomical complexity. The FeTA Challenge 2022 was able to successfully evaluate and advance generalizability of multi-class fetal brain tissue segmentation algorithms for MRI and it continues to benchmark new algorithms. The resulting new methods contribute to improving the analysis of brain development in utero.[1]

## I. INTRODUCTION

IN-UTERO Magnetic Resonance Imaging (MRI) of the fetal brain allows clinicians and researchers to visualize the development of the human brain. The brain development of fetuses can be investigated starting in the second trimester with MRI, and continue up until birth, and can be used in fetuses with both typical neurodevelopment and neurological congenital disorders [1]. It can aid in the future development of clinical perinatal planning tools for early interventions, treatments, and clinical counseling, and can be used to explore complex neurodevelopment of different structures within the brain. Large-scale acquisition and analysis of *in-utero* fetal brain MRI requires collaboration from specialized clinical centers as image cohorts of various patient populations tend to be small at each center. A crucial step of analyzing these MR images involves quantifying the volume and morphology of different anatomical structures in the developing brain, necessitating image segmentation. Manual segmentation is time-intensive, susceptible to variability between observers and centers, making it impractical for extensive collaborative efforts. However, many challenges exist in developing automatic segmentation tools that will work across data from different imaging centers.

[1] This work was supported by the URPP Adaptive Brain Circuits in Development and Learning (AdaBD) project, the Vontobel Foundation, the Anna Müller Grocholski Foundation, the EMDO Foundation and the Prof. Dr Max Cloetta Foundation, the Swiss National Science Foundation (SNSF 320030_184932, 205321–182602), the Austrian Science Fund FWF [P 35189-B, I 3925-B27] and Vienna Science and Technology Fund WWTF [LS20-030]. We acknowledge access to the facilities and expertise of the CIBM Center for Biomedical Imaging, a Swiss research center of excellence founded and supported by Lausanne University Hospital (CHUV), University of Lausanne (UNIL), Ecole polytechnique fédérale de Lausanne (EPFL), University of Geneva (UNIGE) and Geneva University Hospitals (HUG). This work was supported by the NIH (Human Placenta Project—grant 1U01HD087202-01), Wellcome Trust Sir Henry Wellcome Fellowship (201374/Z/16/Z and /B), UKRI FLF (MR/T018119/1), EPSRC (EP/V034537/1), and by core funding from the Wellcome/EPSRC Centre for Medical Engineering [WT203148/Z/16/Z].

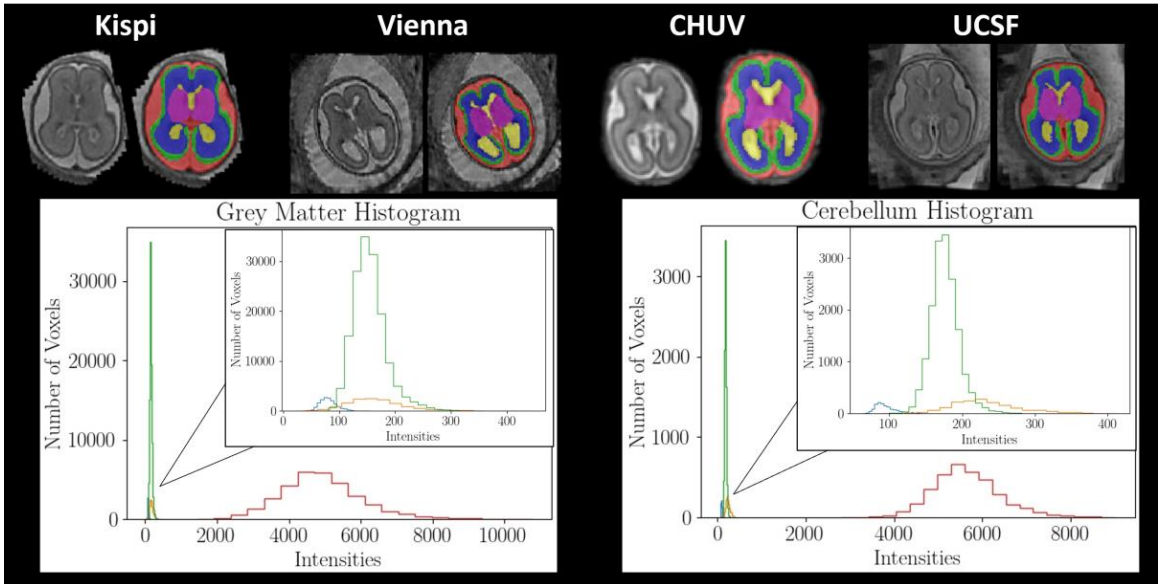

Fig. 1. Sample cases from each institution in the testing dataset. Each case is a normally developing fetal brain from gestational week 22, with a super-resolution quality rating of 'Excellent'. The histograms of the individual labels vary between each institution (green: Kispi, orange: Vienna, blue: CHUV, red: UCSF). The inset is an enlarged view of the first peak to visualize the different histograms of the three institutions.

Existing deep learning-based methods work well when they are tested on similar data to which they were trained on (*i.e.* in-domain data), but struggle when facing testing data that is different from the training data (*i.e.* unseen, or out-of-domain data), such as images acquired at another site, with a different scanner, or with different scanning parameters [2]. Even after careful image processing, classifiers are able to tell the differences between images acquired with different scanners [3]. Efforts to standardize fetal MRI acquisition parameters across different imaging centers or hospitals have been limited, primarily because fetal imaging relies on specialized sequences that are fine-tuned locally. The appearance of MR images is significantly influenced by various factors, including acquisition parameters, magnetic field strength, MRI coil type, overall imaging setup, and the expertise of the technicians performing the image acquisition. These site differences (or domain shifts) have been shown to be very challenging for deep learning algorithms to handle if there is no similar data in the training dataset [2], [4], [5]. Domain generalizability of automatic segmentation algorithms is an urgent need and is attracting increasing attention in the medical imaging field [6]–[10].

In our previous Fetal Tissue Annotation Challenge (FeTA) 2021, we used the first publicly available dataset of fetal brain MRI data to encourage teams to develop automatic fetal brain tissue segmentation methods [11]. However, in this dataset, the training and testing datasets were from the same imaging center. For the FeTA Challenge 2022, we launched a multi-center fetal brain segmentation challenge focused on model generalizability across different imaging centers including two unseen centers.

Here, we describe the multi-center FeTA Challenge 2022 and its organization as well as give an overview of the submitted algorithms and provide a detailed analysis and evaluation of the challenge results. This paper adheres to the transparent reporting guidelines as described in the BIAS method [12]. The aim of the multi-center FeTA Challenge 2022 was to promote the development of domain-robust algorithms for automatically segmenting high-resolution fetal brain MRI reconstructions between 19-35 gestational weeks into seven different classes that would work on data from different imaging centers. The challenge included data from four different imaging centers, further expanding on the FeTA dataset [13]. Two of the centers were included in the training dataset, and all four imaging centers were included in the hidden testing dataset on which the algorithms were evaluated to test on both seen and unseen data. Examples from each site can be seen in Fig. 1. The algorithms were evaluated on the hidden testing dataset. The submitted algorithms were also tested on various subsets of the testing dataset to determine whether they perform better or worse on data from different imaging centers or under different circumstances such as image quality or reconstruction method.

In addition to analyzing the results of the FeTA Challenge 2022, we also proposed to investigate the usage of topology as a new evaluation metric for automatic segmentation algorithms. Given that a key downstream analysis of segmentation is the extraction of surface and surface-based metrics (such as thickness and curvature), computational topology of binary masks (*i.e.* connected component, holes) are important to evaluate. We investigated whether topology errors should be added to current evaluation metrics, , as the metrics chosen for challenges play a significant role in challenge results [14]. This holds particular importance for the analysis of cerebral cortex segmentation, which remains one of the most challenging structures to segment in the developing brain.

The algorithms developed as part of the multi-center FeTA Challenge 2022 have the potential to transform both the clinical and research fetal MRI environment, leading to better antenatal and perinatal tools being developed across hospitals and research institutions around the world.

TABLE I
TRAINING AND TESTING DATASET PROPERTIES FROM ALL IMAGING CENTERS.

| | TESTING DOMAIN | INSTITUTION | SCANNER | N | SUPER-RESOLUTION METHOD | RESOLUTION (MM$^3$) | TR/TE | GESTATIONAL AGE RANGE (WEEKS) |
|---|---|---|---|---|---|---|---|---|
| Training | In Domain | Kispi | GE Signa Discovery MR450/MR750 (1.5T/3T respectively)* | 80 | MIALSRTK (n=40) irtk-simple (n=40) | 0.5x0.5x0.5 | TR: 2000–3500ms, TE: 120ms (minimum) | 20.0-34.8 |
| | | Vienna | Philips Ingenia/Intera (1.5T); Philips Achieva (3T)* | 40 | NiftyMIC** (n=40) | 1.0x1.0x1.0 | TR: 6000-22000ms TE: 80-140ms | 19.3-34.4 |
| Testing | In Domain | Kispi | GE Signa Discovery MR450/MR750 (1.5T/3T respectively)* | 40 | MIALSRTK (n=20) irtk-simple (n=20) | 0.5x0.5x0.5 | TR: 2000–3500ms, TE: 120ms (minimum) | 21.3-34.6 |
| | | Vienna | Philips Ingenia/Intera (1.5T); Philips Achieva (3T)* | 40 | NiftyMIC** (n=40) | 1.0x1.0x1.0 | TR: 6000-22000ms TE: 80-140ms | 18.1-35.0 |
| | Out of Domain | CHUV | Siemens MAGNETOM Aera (1.5T) | 40 | MIALSRTK (n=40) | 1.125x1.125x 1.125 | TR: 1200ms, TE: 90ms | 21.0-35.0 |
| | | UCSF | GE Discovery MR750/MR750W (3T) | 40 | NiftyMIC** (n=40) | 0..8x0.8x0.8 | TR: 2000-3500 ms, TE: 100 ms (minimum) | 20.0-35.1 |

\* The training dataset contained data from both 1.5T and 3T scanners. However, which cases belonged to which scanner were not provided to the participants as it was part of the data anonymization process. Therefore, the breakdown of number of cases per scanner is not provided here.
\*\* When the NiftyMIC algorithm was used, the image included the maternal tissue. The brain mask generated automatically by the algorithm was not used. Therefore, the NiftyMIC cases contained more maternal tissue than the fetal brains reconstructed with the MIALSRTK and irtk-simple algorithms.

## II. METHODS

### A. Challenge Organization

The FeTA Challenge 2022 (feta.grand-challenge.org) was held in conjunction with the Medical Image Computing and Computer Assisted Intervention (MICCAI) 2022. The challenge is a repeated annual event at MICCAI, with a fixed submission deadline. Participants were asked to submit a fully automatic segmentation algorithm that would segment high-resolution fetal brain MRI reconstructions into seven different tissue types: external cerebrospinal fluid (eCSF), grey matter (GM), white matter (WM), ventricles, cerebellum, deep grey matter (deep GM), and brainstem.

In addition to the FeTA training dataset, the participants were able to use additional data for training only if it was publicly available, and were required to document the usage in their algorithm description. Participants were able to modify the provided training data as well. This modification includes the generation of additional data by image synthesis or various data augmentation strategies (for example, using numerical simulations by FaBiAN [15]) as long as everything was documented, and the synthetic data could be made available to challenge organizers upon request.

All teams with valid submissions and who presented their results at MICCAI 2022 are included in this paper. Each team was allowed three co-authors. Participating teams are able to publish their algorithms and results independently after the challenge, but should cite this challenge paper and the data publication paper [13].

The full results were announced at the MICCAI 2022 conference and were published on the challenge website. The top three teams received custom-made FeTA chocolate bars. Participating teams were able to choose whether they wished to make their submission public. The Dockers of all submissions with consent to publicly release can be found here: https://hub.docker.com/u/fetachallenge22. Each team was required to provide a written description of their algorithm, which can be found in the Supplementary Information [16].

Participants were asked to submit a Docker container containing their fully automatic segmentation algorithm to the organizers via email. Members of the organization committee were allowed to participate but were not eligible for awards. The organizers ran the Docker container on the testing datasets using evaluation code available on the challenge website. No multiple submissions were allowed. Resubmissions were only allowed in cases of technical errors with the Docker.

The training dataset was released to participants on June 1, 2022, and the Docker submission deadline was August 3, 2022. The top-performing teams were informed that they were a top-performing team on September 3, 2022 in order for them to prepare a presentation for the day of the challenge. The challenge day was September 18, 2022, where the results were presented at the MICCAI FeTA Challenge 2022 session. For the complete overview of the challenge, see the final challenge proposal [17].

### B. Mission of the Challenge

The mission of the FeTA Challenge 2022 is to encourage and facilitate the development of generalizable automatic multi-class segmentation algorithms that are able to segment the fetal brain into seven different tissue types plus background from

4MRI. To achieve this goal, clinically acquired, anonymized MRI data were used to represent the target cohort, pregnant women who underwent fetal MRI. The accuracy of the fetal brain segmentations was evaluated in the challenge cohort. Fetal brain MRI scans were acquired clinically and reconstructed using super-resolution reconstruction methods. The gestational age, and a label of normal neurodevelopment or pathological neurodevelopment is included for each case in the dataset, and the cases span a gestational age (GA) range of 18-35 weeks.

### C. Challenge Dataset

The challenge dataset consisted of fetal brain MRI reconstructions acquired from four different imaging centers. Data from two centers (University Children's Hospital (Kispi), Medical University of Vienna) is included in the training dataset, and an additional two centers were included in the testing dataset (University Hospital Lausanne (CHUV), University of San Francisco (USCF), for a total of four centers. In this challenge, one case consists of the following: a super-resolution reconstruction of the fetal brain MRI, a manually segmented label map consisting of eight labels (eCSF, GM, WM, ventricles, cerebellum, deep GM, brainstem, background), a gestational age, and the classification of normal or pathological neurodevelopment. The testing dataset was hidden from participants. In total, there were 120 cases in the training dataset and 160 cases in the testing dataset (see overview in Table I). A separate validation dataset was not provided to the participants. The distribution of GAs and the split between normal and pathological neurodevelopment was kept as equal as possible between the two centers included in both the training and testing dataset. For the two unseen imaging centers, a range of gestational ages, pathologies, and normal neurodevelopmental cases were included to mimic the potential real-world usage of automatic segmentation algorithms. Each case in the dataset was manually segmented using the same method. Several annotators with experience in medical imaging were trained to segment different labels (AJako, MK, AA, PM, GG, HJ, CS, KP, AJaka), and then the individual labels were automatically combined. Afterwards, three experts in fetal MRI (KP, CS, AJaka) reviewed and corrected each label map, where each case was reviewed by two of the three experts in a two-step process to minimize error. An analysis of inter-rater agreement of fetal brain manual segmentation can be found in [13]. Exact details of the manual segmentation can be found in the supplementary information of [13].

#### 1) University Children's Hospital Zurich (Kispi) Data
The training and testing data from FeTA 2021 was used in FeTA 2022, and the image acquisition parameters, post-processing steps, and ethical approval information can be found in [18] There were 80 training cases and 40 testing cases. These testing cases are considered in domain, as this site provides both training and testing cases.

#### 2) University of Vienna (Vienna) Data
The data from the Medical University of Vienna was acquired using 1.5 T (Philips Ingenia/Intera, Best, the Netherlands) and 3 T magnets (Philips Achieva, Best, the Netherlands), without the use of maternal or fetal sedation. All acquisitions were performed using a cardiac coil. For each case, at least 3 T2-weighted single-shot, fast spin echo (SSFSE) sequences (TE=80-140ms, TR=6000-22000ms) in 3 orthogonal (axial, coronal, sagittal) planes with reference to the fetal brain stem axis and/or the axis of the corpus callosum were acquired. Overall, slice thickness was between 3mm and 5mm (gap 0.3-1mm), pixel spacing 0.65-1.17mm, acquisition time between 13.46 and 41.19 seconds.

The preprocessing pipeline [19] consists of a data denoising step [20], followed by an in-plane super resolution [21] and automatic brain masking step [22] and concludes with a single 0.5 mm isotropic slice-wise motion correction and volumetric super-resolution reconstruction [22]. Subsequently, the resulting volumes are rigidly aligned to a common reference space [23].

Fetal MRI cases were provided by the Medical University of Vienna. The data has been acquired as part of a retrospective single-center study and has been anonymized and approved by the ethics review board and data clearing department at the Medical University of Vienna, responsible for validating data privacy and sharing regulation compliance. There were 40 training cases 40 testing cases included in the FeTA Challenge 2022 from this site. As with the Kispi Data, this is considered an in-domain testing set, as this site provides both training and testing data.

#### 3) Lausanne University Hospital (CHUV) Data
The data from CHUV was acquired at 1.5T (MAGNETOM Aera, Siemens Healthcare, Erlangen, Germany), without the use of maternal or fetal sedation. Acquisitions were performed with an 18-channel body coil and a 32-channel spine coil. Images were acquired using T2-weighted (T2W) Half-Fourier Acquisition Single-shot Turbo spin Echo (HASTE) sequences in the three orthogonal orientations (axial, sagittal, coronal); usually at least two acquisitions were performed in each orientation., TR/TE, 1200ms/90ms; flip angle, 6/23 90°; echo train length, 224; echo spacing, 4.08ms; field-of-view, 360 × 360mm2 ; voxel size, 1.13 × 1.13 × 3.00mm3 ; inter-slice gap, 10%, acquisition time between 26 to 36 seconds.

For each subject, the scans were manually reviewed and the good quality scans were chosen for super-resolution reconstruction, creating a 3D SR volume of brain morphology [24]. Each case was zero-padded to 256x256x256 and reoriented to a standard viewing plane. Mothers of all other fetuses included in the current work were scanned as part of their routine clinical care. Data was retrospectively collected from acquisitions done between January 2013 to April 2021. All images were anonymized. This dataset is part of a larger research protocol approved by the ethics committee of the Canton de Vaud (decision number CER-VD 2021-00124) for re-use of their data for research purposes and approval for the release of an anonymous dataset for non-medical reproducible research and open science purposes. As no training cases were included from this site, the 40 testing cases are considered out of domain.

*4)* University of California San Francisco (UCSF) Data

The data from UCSF was acquired using 3T GE Discovery MR750 or MR750W (wide bore) without the use of maternal or fetal sedation. Acquisitions were performed using a 32 channel GE cardiac coil. At least 3 T2-weighted ssFSE sequences were acquired with one scan per orientation (sagittal, axial, coronal) with the following parameters: 240 mm FOV with 512x512 matrix gives in plane resolution of ~0.5x0.5 mm with 3 mm slice thickness. TR is 2000-3500 ms, TE > 100 ms, 90 deg flip angle.

For each subject, the scans were manually reviewed and the good quality scans were chosen for super-resolution reconstruction, creating a 3D SR volume of brain morphology [22]. Each case was zero-padded to 256x256x256 and reoriented to a standard viewing plane.

Fetal MRI was acquired during routine clinical care with institutional review board approval for anonymized retrospective analysis by the FeTA team (IRB 21-35930). As no training cases were included from this site, the 40 testing cases are considered out of domain.

### D. Evaluation Metrics

Three complementary types of evaluation metrics were used to compute the rankings. The overlap is quantified with the dice similarity coefficient (DSC) [25]. The similarity between the two volumes is quantified with the volume similarly measure [25]. The contours are evaluated with a boundary-distance-based metric: the 95th percentile of the Hausdorff distance (HD95) (https://github.com/deepmind/surface-distance). As the task is a segmentation task, the DSC was chosen, as it is the most popular segmentation metric. However, we would like not just an overlap metric, but we are also interested in the shape and volume, as shape and volume are often used as clinical biomarkers. Therefore, we included the HD95 (shape) and VS (volume). The final rankings will take all three metrics into account.

### E. Ranking

The ranking method was the same as in FeTA 2021 [18]. Each of the participating teams was ranked based on each evaluation metric, and then the final rankings combined the rankings from all of the metrics (DSC, HD95, VS) for the complete dataset (both in and out of domain imaging site). The DSC, HD95, and VS were calculated for each label within each of the corresponding predicted label maps of the fetal brain volumes in the complete testing dataset. The mean and standard deviation of each label for all test cases was calculated, and the participating algorithms were ranked from low to high (HD95), where the lowest score received the highest scoring rank (best), and from high to low (DSC, VS), where the highest value received highest scoring rank (best) based on the calculated mean across all labels and test cases. If there were missing results, the worst possible value is used. For example, if a label does not exist in the new segmentation label map but is present in the ground truth (GT) label map, it will receive a DSC and VS score of 0, and the HD95 score will be double the max value of the other algorithms submitted. This ranking procedure was developed to take three different metric types equally into account.

Finally, the results of the challenge were run through the ChallengeR toolkit, specifically designed to calculate and display imaging challenge results [26].

Additional rankings were created based on the in-domain and out-of-domain imaging centers, cases with and without neurological pathologies, and image reconstruction quality (Excellent, Good, Poor). These additional rankings were not part of the determination of the winner of the challenge but were presented at the FeTA Challenge 2022 event.

### F. Topology Analysis

In addition to the rankings mentioned in the previous section, we assessed the topology correctness as an evaluation metric of the predicted label maps. Topology defines the properties of an object that are preserved through deformation [27]. Given binary maps (tissue labels), computational topology relies on connectivity of a voxel to its neighbours to quantify the number of connected components, holes, or cavities. Topology is relevant for exploring brain tissue segmentations as topology correctness is needed to quantify biomarkers important for brain development such as cortical thickness and gyration. However, fetal cortical segmentations are often discontinuous [28]–[30] but surprisingly topology correctness of predicted segmentations is rarely reported [31], [32]. Here, we propose a topology-integrative ranking of the methods: the global BNE topology ranking.

To quantitatively compare the topology of each segmented structure, we assessed the error of the topological invariant Betti numbers. The *k*-dimensional Betti numbers ($BN_k$) count the topological structures in each dimension *k*. More specifically, $BN_0, BN_1, BN_2$ represent the number of connected components, the number of holes and the number of cavities in the 3D binary object respectively. We define the *k*-dimensional Betti number error ($BNE_k$) as the absolute difference of the GT expected value and the prediction measure. $BNE_k$ are difference metrics that must be minimized. The GT expected values are as follows: the $BN_1 = 0$ and $BN_2 = 0$ for all brain tissue labels. For eCSF, WM, ventricles, cerebellum, dGM, and brainstem, $BN_0 = 1$, and for GM, $BN_0 = 2$.

When performing the evaluation of the predicted label maps, when there is an absence of segmentation for a tissue, it was attributed twice the value of the worst performing segmentation of the same label over all submissions, in line with how missing data was handled with the HD95 evaluation metric.

Once topology was quantified, we also computed the ranking of methods for each $BNE_k$ with Challenge R toolkit [26]. We inferred a global topology ranking $BNE$ as the ranking of the sum of all three $BNE_k$ rankings.

## III. RESULTS

### A. Challenge Submissions

There were 17 submissions from 16 different teams to the FeTA Challenge 2022. One team submitted two algorithms, but they were determined to be substantially different methodologies and as such was allowed. Each team submitted a written description of their algorithm, with can be found in the Supplementary Information [16]. Two teams used only one institution's dataset rather than the complete training dataset (deepsynth, ajoshiusc). All other teams used the complete training dataset. Seven teams used additional publicly available datasets for pre-training or training (FIT_1, FMRSK, symsense, FIT_2, DBC Pasteur, fudan_zmic, deepsynth).

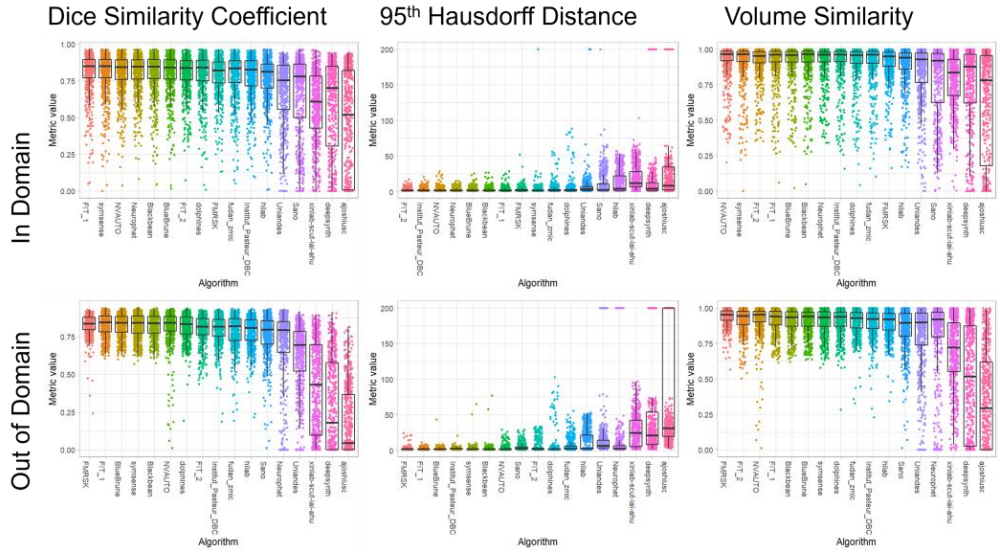

Fig. 2. In Domain and Out of Domain evaluation metrics by algorithm. In both in and out of domain, as well as for all three evaluation metrics (Dice Similarity Coefficient, 95th Hausdorff Distance, Volume Similarity), the results plateau for the first 10 teams, after which a drop off is observed. The ranking of the teams has changed between the In Domain and Out of Domain metrics.

All submitted models relied on deep learning. Only three teams used 2D networks (fudan_zmic, DBC Pasteur, ajoshiusc), the remainder of the teams used 3D networks. All teams used PyTorch, or Pytorch-based solutions (such as nnUnet [33] or MONAI [34]) for their network. Many teams used a two-step strategy for segmentation (often classified as 'coarse-to-fine'). This often involved first segmenting the brain from the outlying maternal tissue, and then segmenting the fetal brain into different tissues. Each algorithm is summarized in further detail in Table III. Institutional ranking differences in the submissions can be found in Fig. 3.

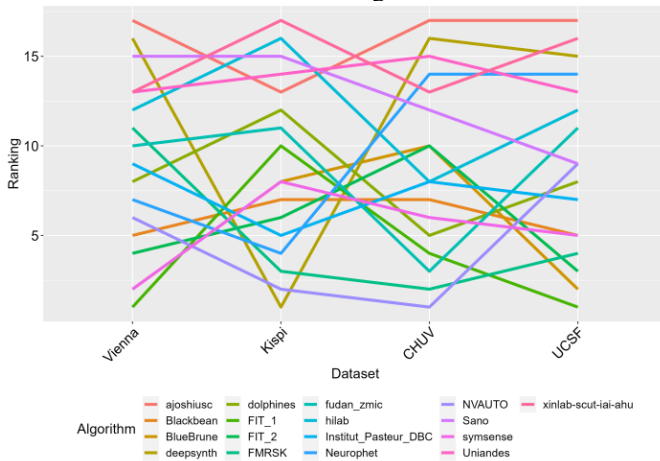

Fig. 3. Rankings of participating teams separated by each institutions test dataset. In-domain institutions: Vienna, Kispi; Out of domain institutions: CHUV, UCSF

### B. In-domain results

In-domain evaluation is defined based on the performance on the subset of data including the Kispi and Vienna data, as data from these two imaging centers were represented in the training dataset available to the participants. A summary of the in-domain evaluation metrics for all teams can be seen in the top row of Fig. 2. We report two aspects of the in-domain evaluation results. Firstly, we present in-domain team rankings and an in-depth evaluation of the FeTA Challenge 2022 results. Secondly, we cross-reference these rankings with the outcomes achieved in the FeTA Challenge 2021 [18]. Notably, the Kispi data included in the FeTA2022 is identical to the FeTA 2021 training dataset (80 cases).

In the overall ranking of the in-domain dataset, the top three submissions were *NVAUTO*, *FIT_2* and *FIT_1*. Specifically, *FIT_1* (0.8052), *symsense* (0.8047), and *NVAUTO* (0.8042) were the top three teams according to the DSC. The top three submissions according to the HD95 were *FIT_2* (2.31mm), *Institut_Pasteur_DBC* (2.40mm), and *NVAUTO* (2.46mm). The top three submissions according to the VS were *NVAUTO* (0.914), *symsense* (0.910) and *FIT_2* (0.910). It is worth noting that no statistically significant differences were found in the rankings for the achieved DSC scores in the first four teams, (*FIT_1*, *symsense*, *NVAUTO*, *Neurophet*). In the HD95, the first ranked submission, *FIT_2*, was significantly better performing than the second ranked (*Institut_Pasteur_DBC*), while the top two teams in VS (*NVAUTO* and *symsense*) were not significantly different. Further details about the individual rankings are shown in Fig. 2. Similar to the FeTA 2021 Challenge, a performance plateau was observed in the DSC scores, with approximately the first 12 teams achieving very similar DSC scores (DSC range for the top 12 teams: 0.765 – 0.805), with a large drop off in scores in the last five submissions (DSC range for the last 5 teams: 0.455 – 0.684). A similar trend was observed for the mean HD95 (highest ranked 9 submissions: 2.31 to 2.83 mm, lowest ranked 8 submissions: 3.5 to 41 mm) and for the mean VS scores (highest ranked 11 submissions: 0.902 – 0.914, lowest ranked 6 submissions: 0.611 - 0.880).

Not all anatomical structures were segmented equally well, which is reflected by the heterogeneity of mean DSC, HD95 and VS scores obtained in the in-domain evaluation. The white matter and ventricles were the structures most successfully segmented. Here, we outline the DSC scores of the labels in detail. The mean DSC for the top three submissions for the white matter were 0.885 (*FIT_1*), 0.883 (*symsense*) and 0.882 (*Blackbean*), and the ventricles were 0.889 (*NVAUTO*), 0.889 (*symsense*) and 0.888 (*FIT_1*). On the other hand, the cortex was segmented rather poorly, as the mean DSC for the top three submissions were 0.726 (*FIT_1*), 0.725 (*NVAUTO*) and 0.724 (*Neurophet*). The external CSF spaces, which neighbor the cortex, were similarly poorly segmented.

Compared to the FeTA Challenge 2021 results, segmentation accuracy improved marginally. The highest DSC in the FeTA Challenge 2022 in-domain evaluations was 0.805, while it was 0.786 in 2021. The lowest HD95 in the FeTA2022 in-domain evaluation was 2.31 mm, while it was 14 voxels in 2021. These two metrics are not directly comparable due to the change in evaluation tool and unit between the years, as the tool used in FeTA 2021 was not ideal when outliers were present. The highest average VS in the FeTA Challenge 2022 was 0.914, while it was 0.885 in 2021. In-domain, the per-label comparisons yielded similar results: the cortex and the eCSF being the most difficult to segment, while the WM and the ventricles the best performing. There were two teams who submitted to both FeTA 2021 and FeTA 2022 who ranked very well in the in-domain evaluation: *NVAUTO* and *Neurophet*. *NVAUTO* maintained a top in domain ranking in both years in all three evaluation criteria (2021: DSC 1st place, HD95 1st place, VS 2nd place, 2022: DSC 3rd place, HD95 3rd place, VS 1st place), as did *Neurophet* (2021: DSC 3rd place, VS 5th place, 2022: DSC 4th place, HD95 4th place).

### C. Inter-site Generalizability assessment: out-of-domain performance

Here, we evaluate the performance of the submissions on unseen datasets (i.e. on data that was not present in the training dataset). Therefore, we present the out-of-domain (OOD) performance rankings using the CHUV and the UCSF testing data subset and compare them with the in-domain results. A summary of the OOD evaluation metrics for all teams can be seen in the bottom row of Fig. 2. Some submissions demonstrate equivalent performance for both the in-domain and OOD subsets such as *FIT_1* (ranked 3rd in-domain and 2nd OOD), *Symsense* (ranking 4th for both in-domain and OOD), or *Dolphins* (ranking 9th for both in-domain and OOD). Interestingly, some methods rank better in the OOD subset, such as *BlueBrune*, which rises from 6th place in-domain to 3rd place OOD, or *Blackbean* who rises from 7th rank in-domain to 4th OOD. However, some models drop considerably in performance such as *FIT_2* (from 2nd to 7th), *NVAUTO* (from 3rd to 6th, performing poorly in many OOD cases, see Fig. 3 bottom row) or *Neurophet* (from 4th to 13th). This indicates that the domain shift present in data from different imaging centers can drastically degrade model performance when being deployed in heterogenous clinical datasets.

Overall, the median performance metrics in the OOD setting remain equivalent to the in-domain, with many of the models attaining a plateau of performance around 0.80, 2.5, 0.90 in DSC, HD95 and VS respectively. However, the median of the worst performing methods dropped by a large amount (dropping to approximately 0 for DSC, or 0.25 for VS) while in-domain median performance never reaches such low levels (always above 0.50 and 0.75 for DSC and VS respectively for all methods).

Not all brain tissue labels are equivalent when comparing in-domain and OOD results. Class-wise performance (see Supplementary Information, Section 12 [16]) indicates that major drops of performance occur in ventricles (in DSC, HD95, VS), and GM and WM volume (in VS). The achieved performance by top ranking algorithms in the other tissues (eCSF, deepGM, cerebellum, brainstem) were even slightly higher OOD than in domain (eg. DSC range of 0.83 to 0.36 OOD while 0.76 to 0.04 in domain).

### D. Global ranking

The global ranking is the ranking as defined by using the complete testing dataset from all four imaging centers. The global ranking is the official ranking which determined the winners of the FeTA Challenge 2022.

Examples of results from the top 5 teams can be found in Fig. 4. The team rankings of each evaluation metric can be seen in Fig. 5, and the rankings based on the different labels can be found in Fig. 6. The final rankings can be found in Table IV.

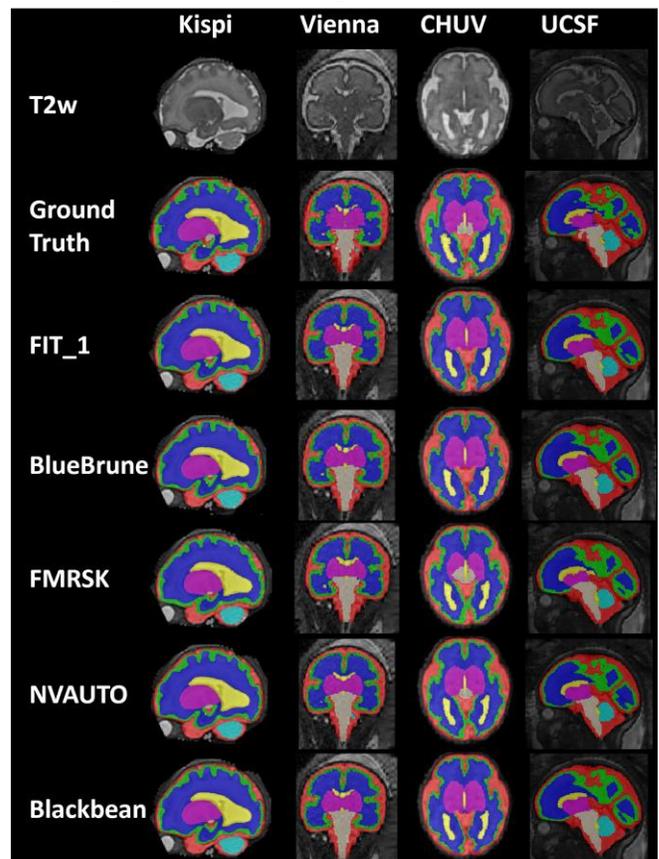

Fig. 4. Examples of the automatic labels created by the top 5 teams for each of the institutions (T2w: T2-weighted fetal brain reconstruction).

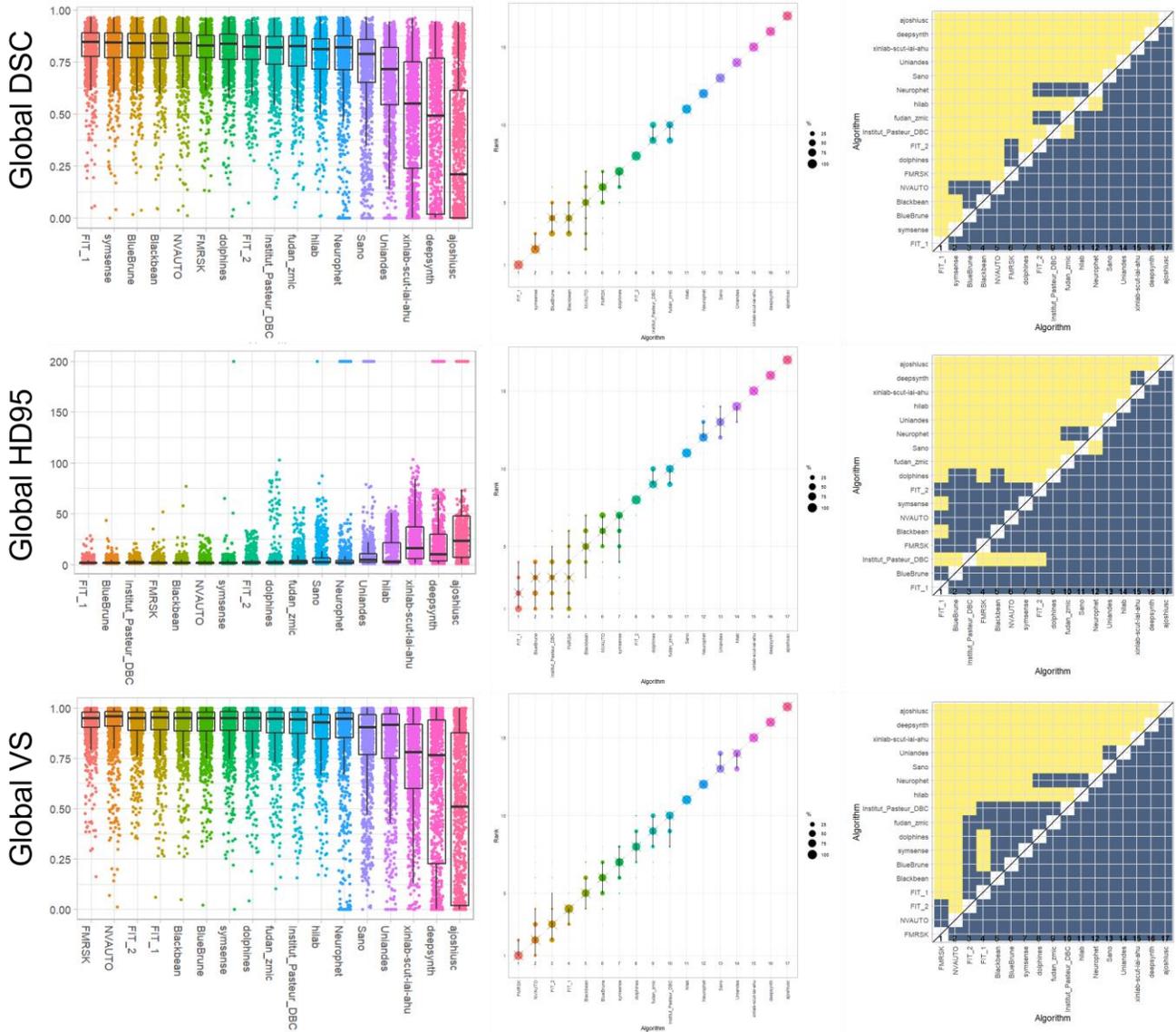

Fig. 5. Rankings of participating teams for each metric from top to bottom (left to right). Left column: Middle Column: Right Column: DSC: Dice Similarity Coefficient: HD95: 95[th] Hausdorff Distance: VS: Volume Similarity

The top three teams are *FIT_1*, *Bluebrune*, and *FMRSK* (with *Bluebrune* and *FMRSK* tied for second). *FIT_1* maintained a top 5 ranking across each of the labels, while the rankings were much more variable for all other teams across the different brain tissues. A plateau in performance of the top 10-12 teams was observed, in line with the in-domain and out of domain results. The top three DSC scores were from teams *FIT_1* (0.816), *symsense* (0.813), and *Bluebrune* (0.812). The top three HD95 scores were from *FIT_1* (2.35mm), *Bluebrune* (2.38mm), and *Institute_Pasteur_DBC* (2.39mm). The top three three VS scores were from team *FMRSK* (0.920), *NVAUTO* (0.915), and *FIT_2* (0.913).

In order to investigate factors which may have influenced the ratings, we looked at rankings based on quality ratings of the testing dataset (Excellent=3, Good=2, Poor=1, median rating by 3 experienced individuals), normal and pathological brains, as well as rankings based on the super-resolution reconstruction algorithm used (NiftyMIC, mial-srtk, irtk-simple).

For the excellent quality fetal brain reconstructions, the top three teams were *FIT_1*, *FMRSK*, and 4 teams tied for 3[rd] (*symsense, NVAUTO, Blackbean, BlueBrune*). The 'Good Quality' top three teams were *FIT_1*, *FMRSK*, and *NVAUTO*, and 'Low Quality' were *BlueBrune*, *NVAUTO*, and *FIT_1*. The top three teams for fetal brains with the normal classification, were *FMRSK*, *FIT_1*, and *NVAUTO* and for pathology were *FIT_1*, *BlueBrune*, and *FMRSK*. The top three teams for fetal brains reconstructed with the irtk-simple algorithm [35] were *deepsynth*, *FMRKS*, and *ajoshiusc*; with mial-srtk algorithm [24] were *FMRSK*, *NVAUTO*, and *fudan-zmic*; and with the NiftyMIC algorithm [22] were *FIT_1*, *BlueBrune,* and *Blackbean*. When separating the rankings based on individual labels, *BlueBrune* was the top ranking team for the eCSF, *NVAUTO* ranked first for the GM, *FMRSK* ranked first for the brainstem, and *FIT_1* was the top team for the remaining labels (WM, ventricles, cerebellum, deepGM). A complete overview of the rankings per label can be found in Fig. 6 as well as in the Appendix.

TABLE III
FETA 2022 TEAM OVERVIEW

| Rank | Team Name | Architecture | Training Strategy | Loss Function | Post-processing | Augmentation | External Datasets |
|---|---|---|---|---|---|---|---|
| 1 | FeTA-ICL-TUM (FIT) – nnUnet (FIT_1) | nnU-Net | Ensemble of 5 different models | CE and soft Dice | Ensemble + rule-based denoising autoencoder post-processing | (i): Default nnUnet augmentation; (ii): (i) + random bias field; (iii): (i) + style augmentation, random bias field; (iv): (i) + photometric augmentation; (v): (i) + motion artifact | ImageNET backbone |
| 2 | BlueBrune* | nnU-Net | Data Split: 80/20; (i) tissue segmentation network; (ii) domain adversarial approach | CE and dice | ensemble (2 models) | default nnU-Net augmentation | No |
| 2 | FMRSK* | (i) 3D U-Net (ii) Attention U-Net (MONAI) | (i) brain extraction; (ii) tissue segmentation | Dice and CE (MONAI) | Average prediction of 2 models | motion artifact, MR spike, Bias field, affine transform, noise, blurring, gamma, random intensity shift | 19 dHCP neonates, Spina bifida atlas |
| 4 | NVAUTO | SegResNet | 5-folds CV | Dice Focal from MONAI | ensemble on average prediction (15 models) | normalize to zero mean and unit standard deviation | No |
| 5 | Blackbean | (i) nnU-Net, (ii) ViT-Adaptor | Data Split: 100/0; (i) two models, (ii)Test-time augmentation | Dice and BCE | Ensemble on test-time augmented (softmax mean) | default nnU-Net augmentation | No |
| 6 | symsense | nnU-Net | 5-folds CV | modified Generalized Dice and CE | None | Default nnUnet augmentation, brightness transform, contrast transform, zoom, warping, GIN-IPA | 40 early neonatal dHCP |
| 7 | FeTA-ICL-TUM (FIT) – SWINUNETR (FIT_2) | (i) Synthstrip, (ii) Swin UNETR (MONAI) | (i) skull stripping; (ii) tissue segmentation | weighted CE and soft Dice | Resampling to original image | flipping, rotation, affine + elastic transformation, noise, blur, gamma, ghosting, spike, motion, bias, blur, anisotropy | neonate subjects of the dHCP |
| 8 | DBC Pasteur | U-Net | Ensemble of 3 2D networks (ax/sag/cor) Data split 80/20% | CE | Ensemble (3 models) majority vote | Noise, blur, 2D rotations + translations + flip, zoom | Atlas Gholipour + Atlas Serag |
| 9 | Dolphins | (i) 3DResUNet (coarse), (ii) nnUNet (fine) | 5-folds CV, keep the one with best validation scores | BCE+Dic | None | Flipping, random gamma | No |
| 10 | fudan_zmic | BayeSeg (based on nnU-Net) | 5-folds CV | CE, Dice, and weighted variational | Ensemble (5 folds cv) | cardiac cutmix augmentation to enrich the background, and nnU-Net augmentations | ACDC dataset for cutmix |
| 11 | hilab | (i) nnU-Net, (ii) residual 3D U-Net | (i) coarse multiclass model, (ii) seven fine single class models | CE and Dice loss | Ensemble of stage (ii). | rotation and scaling, Gaussian noise and blur, brightness and contrast adjustment, simulation of low resolution, gamma augmentation, mirroring | No |
| 11 | Neurophet | 3D U-Net | 3 models trained; probability-based sampling method to focus network | Dice, CE (custom weight) | All 3 models used, non-zero voxels measured, volumes compared | spatial (horizontal flip, rotation, affine transform) and intensity (gaussian blur) | No |
| 13 | Sano | Swin UNETR | 5-folds CV | CE and Dice | Ensemble learning (5 models) | cropping, random zoom, random rotation, random gaussian noise, random adjust contrast, random flip on each axis | No |
| 14 | Uniandes | ROG (from MSD) | 2-folds CV | Dice and CE | closure and opening of grays in segmentation map with structuring element | Spatial Transform (random rotation and scaling), Mirror Transform and gamma correction. | No |
| 15 | xinlab-scut-iai-ahu | (i) DynUNet, (ii) Swin Transformer, (iii) SwinUNETR | 5-folds CV; (i) brain extraction, (ii) Domain generalization stage, (iii) tissue segmentation | (i) Dice, (ii) Contrastive, CE, L2 (iii) Dice | argmax operated results of sliding window patches with 0.5 overlap | orientation change, spacing, cropping, flipping, rotation | No |
| 16 | deepsynth | U-Net, SynthSeg | Training on synthetic, fine tuning on real T2w (dHCP, FeTA) | Mean dice score | None | Synthetic MRI from dHCP label-maps with Synthseg | 80 youngest dhCP subjects |
| 17 | ajoshiusc | R50-ViT: combo of ResNet-50 and ViT, transunet | Data Split: 75/5 subjects. | CE, robust CE based on beta divergence | None | None | No |

BCE: Binary Cross-Entropy, CE: Cross-Entropy, CV: cross-validation, dHCP: developing Human Connectome Project, MSD: Medical Segmentation Decathlon

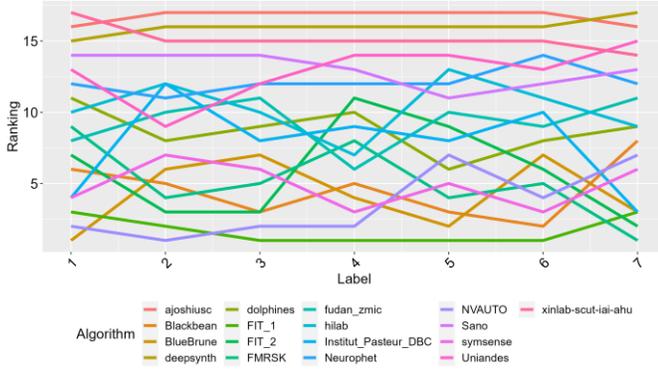

Fig. 6 - Rankings of participating teams separated by label

### D. Topological Analysis Results

Table V (A) presents the topology-integrative ranking (TIR) of the submissions for each dimension $k \in \{0,1,2\}$ and the global topological rating (BNE). The topology rankings seem similar across the three $k$-dimensional BNEs, with a maximum rank difference of less than three with one exception: *FMRSK* presents a relatively big delta in its BNE rankings of dimension 1 (rank=4) and 2 (rank=13). We hypothesize that such inter-dimension variation may come from tissue-specific errors. Interestingly, *hilab*, which does not perform well in $BNE_0$ (rank=10) and $BNE1$ (rank=11), is the best performing submission in $BNE_2$ (rank=8). Nonetheless, the good $BNE_2$ performance is not sufficient to pass on to the global BNE ranking.

Changes in the global ranking (see Table V(B)) are small: maximum one rank difference except for *Blackbean*, which goes from rank 5 in the global FeTA ranking to rank 3 using the TIR. The winner and second submissions remain the same.

Table VI presents the global topology BNE ranking of the submissions per tissue class. The TIR of the individual tissues vary quite a bit. For instance, *hilab* ranks first for the eCSF, but ranks 13 in the WM.

Apart from the unquestionable top 2 teams, *FIT_1* and *BlueBrune*, only *Blackbean* and *Dolphins* manage to rank in the upper half of the table for all tissue class. Specifically, the average tissue TIR of *Blackbean* is 3.3, while *FMRSK* ranks on average 9.1.

## IV. DISCUSSION AND CONCLUSION

The practical value of MRI segmentation methods in clinical settings depends on their ability to effectively generalize to previously unseen data. When imaging the developing human brain in vivo, the utilization of various postprocessing methods, including image reconstruction, MRI acquisition, and related acquisition settings, may increase differences between imaging sites. Additionally, the overall image quality tends to be lower in comparison to MRI scans of the adult human brain, leading to less distinct delineation of anatomical structures. Our results have shown that generalizability across multiple sites remains a challenge for MRI segmentation, but resources such as multi-site datasets have the potential to improve the performance of such methods. For example, the top scoring team of the Kispi dataset did not train on the second available dataset, and performed poorly on the other three datasets, leading us to believe that this network was overfitted. For some methods (but not all), there seemed to be a preference for a given super-resolution method in the rankings. The winning team (*FIT_1*) ranked first in the Vienna and UCSF datasets, which were both reconstructed with the NiftyMIC super-resolution algorithm.

TABLE IV
FINAL RANKINGS AND RESULTS OF THE FeTA 2022 CHALLENGE. IN ADDITION, THE SEPARATE RANKINGS FOR THE IN DOMAIN DATASETS (KISPI, VIENNA) AND THE OUT OF DOMAIN DATASETS (CHUV, UCSF) ARE SHOWN.

| GLOBAL RANKING | TEAM NAME | GLOBAL AVERAGE DSC | GLOBAL AVERAGE HD95 | GLOBAL AVERAGE VS | IN-DOMAIN RANKING | OUT-OF-DOMAIN RANKING |
|---|---|---|---|---|---|---|
| 1 | FIT_1 | 0.816 ± 0.11 | 2.347 ± 2.51 | 0.910 ± 0.12 | 3 | 2 |
| 2* | Bluebrune | 0.812 ± 0.11 | 2.377 ± 2.55 | 0.908 ± 0.11 | 6 | 3 |
| 2* | FMRSK | 0.808 ± 0.11 | 2.395 ± 2.94 | 0.920 ± 0.10 | 9* | 1 |
| 4 | NVAUTO | 0.810 ± 0.13 | 2.608 ± 3.30 | 0.915 ± 0.12 | 1 | 4* |
| 5 | Blackbean | 0.812 ± 0.12 | 2.506 ± 3.66 | 0.909 ± 0.11 | 7 | 4* |
| 6 | Symsense | 0.813 ± 0.12 | 2.660 ± 6.81 | 0.907 ± 0.12 | 4 | 4* |
| 7 | FIT_2 | 0.798 ± 0.11 | 3.421 ± 5.51 | 0.913 ± 0.10 | 2 | 7 |
| 8 | Institute Pasteur (DBC) | 0.789 ± 0.13 | 2.387 ± 1.97 | 0.901 ± 0.12 | 8 | 8 |
| 9 | Dolphins | 0.806 ± 0.12 | 4.521 ± 11.87 | 0.905 ± 0.12 | 9* | 9 |
| 10 | Fudan_zmic | 0.788 ± 0.13 | 4.720 ± 7.58 | 0.903 ± 0.12 | 11 | 10 |
| 11 | Hilab | 0.774 ± 0.13 | 13.008 ± 14.84 | 0.887 ± 012 | 12* | 12 |
| 12 | Neurophet | 0.739 ± 0.22 | 10.288 ± 35.54 | 0.844 ± 0.25 | 5 | 14 |
| 13 | Sano | 0.709 ± 0.22 | 7.171 ± 12.76 | 0.817 ± 0.23 | 14 | 11 |
| 14 | Uniandes | 0.652 ± 0.22 | 11.366 ± 27.61 | 0.814 ± 0.23 | 12* | 13 |
| 15 | Xinlab-scut | 0.494 ± 0.29 | 23.150 ± 20.66 | 0.731 ± 0.22 | 15 | 15 |
| 16 | Deepsynth | 0.433 ± 0.34 | 36.653 ± 62.13 | 0.604 ± 0.38 | 16 | 16 |
| 17 | Ajoshiusc | 0.319 ± 0.33 | 56.598 ± 75.01 | 0.480 ± 0.38 | 17 | 17 |

*Tied

TABLE V
TOPOLOGY (A) AND GLOBAL (B) RANKINGS OF THE SUBMISSIONS. (A) BETTI NUMBER ERRORS (BNE) PER DIMENSION AND OVERALL. (B) COMPARISON OF THE FETA CHALLENGE 2022 RANKING AND THE TOPOLOGY-INTEGRATIVE RANKING (TIR). TOP 3 SUBMISSIONS ARE SHOWN IN BOLD

|  | (A) TOPOLOGY | | | | (B) GLOBAL | |
| --- | --- | --- | --- | --- | --- | --- |
| TEAM NAME | K-DIM BNE | | | BNE | TIR | FETA |
|  | BNE0 | BNE1 | BNE2 | | | |
| ajoshiusc | 16 | 15 | 16 | 16 | 17 | 17 |
| Blackbean | 3 | 3 | 4 | 3 | **3** | 5 |
| BlueBrune | 2 | 2 | 3 | 2 | **2** | **2** |
| Deepsynth | 17 | 16 | 17 | 17 | 16 | 16 |
| Dolphins | 5 | 7 | 7 | 5 | 8 | 9 |
| FIT_1 | 1 | 1 | 2 | 1 | **1** | **1** |
| FIT_2 | 7 | 8 | 5 | 6 | 7 | 7 |
| FMRSK | 9 | 4 | 13 | 9 | 4 | **3** |
| Fudan_zmic | 8 | 10 | 9 | 10 | 10 | 10 |
| Hilab | 10 | 11 | 1 | 8 | 11 | 11 |
| Institut_Pasteur_DBC | 11 | 12 | 11 | 12 | 9 | 8 |
| Neurophet | 14 | 14 | 15 | 14 | 12 | 12 |
| NVAUTO | 6 | 6 | 8 | 7 | 5 | 4 |
| Sano | 13 | 13 | 12 | 13 | 13 | 13 |
| symsense | 4 | 5 | 6 | 4 | 6 | 6 |
| Uniandes | 12 | 9 | 10 | 11 | 14 | 14 |
| Xinlab-scut-iai-ahu | 15 | 17 | 14 | 15 | 15 | 15 |

Teams *NVAUTO* and *FMRSK* performed similarly well on the CHUV and Kispi dataset, which both included reconstructions performed using the MIALSRTK super-resolution method.

Our findings further indicate that image augmentation stands as a critical factor in achieving domain generalization. Traditional techniques like affine transformations and contrast adjustments have demonstrated their effectiveness within established segmentation frameworks, including nn-UNet. However, the optimum choice of augmentation techniques remains unclear. As highlighted in Table III, it is noteworthy that the leading teams utilized random bias field and motion artifact augmentations. A deeper analysis of the top-performing teams approaches reveals that style and photometric augmentations (contrast, blur, sharpness, etc.), known for their ability to induce significant intensity distribution variations, could be pivotal for enhancing model generalization. This concept aligns with research into generalizable cardiac structure segmentation [36], [37]. Importantly, a potential trade-off between in-domain and out-of-domain data generalization should be acknowledged [38]. For instance, the *NVAUTO* team, which scored first place in the in-domain data performance, did not use any specialized domain generalization techniques, yet fell to fourth for out-of-domain data. Conversely, *FIT_1*, initially third for in-domain data, rose to first in the overall ranking, underscoring the indispensability of domain generalization in the development of robust image segmentation models.

Interestingly, the performance metrics of the OOD images for some algorithms were not worse than the metrics for the in-domain images. In Fig. 2 the range of evaluation metrics for the in-domain results is much larger than the OOD results. This is primarily driven by the quality of the fetal brain reconstructions, as the average quality ratings of the OOD datasets (UCSF: 2.33; CHUV: 2.35) were higher than the average in-domain dataset quality ratings (Kispi: 2.18, Vienna: 1.95). Therefore, the quality of the fetal brain reconstructions plays a large role in the success of the automatic segmentations.

The top team (*FIT_1*) performed extraordinarily well across all labels, ranking first for four out of seven labels. The rankings of the other top teams were not as stable when looking at the individual fetal brain tissue labels, and no pattern can be found. As in the global rankings, the OOD volumes typically had better evaluation metrics for each individual label, apart from the ventricles and brainstem. It is uncertain why these two labels trend differently when compared to the remaining labels.

There are certain algorithm and training strategy trends that have not changed from FeTA 2021. All entries used deep learning and primarily a 3D architecture. nnUnet remains a popular and effective tool for medical image segmentation, and the most popular loss functions were the Dice Similarity Coefficient loss and cross-entropy loss, or a combination of the two. Extensive data augmentation strategies are an integral aspect of training, and the addition of external datasets did not necessarily lead to better results (*deepsynth* included dHCP subjects and ranked 16[th]).

TABLE VI
TOPOLOGY (BNE) RANKING OF THE SUBMISSIONS PER TISSUE CLASS AND ON AVERAGE.

| TEAM NAME | CSF | GM | WM | VENTRICLES | CEREBELLUM | DGM | BRAINSTEM | AVERAGE |
| --- | --- | --- | --- | --- | --- | --- | --- | --- |
| Ajoshiusc | 10 | 15 | 5 | 17 | 16 | 17 | 14 | 13.4 |
| Blackbean | 4 | 4 | 1 | 2 | 4 | 6 | 2 | 3.3 |
| BlueBrune | 5 | 7 | 2 | 3 | 1 | 1 | 3 | 3.1 |
| Deepsynth | 14 | 16 | 7 | 15 | 17 | 16 | 17 | 14.6 |
| Dolphins | 7 | 8 | 3 | 8 | 2 | 5 | 6 | 5.6 |
| FIT (nnUNet) | 2 | 5 | 4 | 1 | 3 | 2 | 1 | 2.6 |
| FIT (SWINUNETR) | 3 | 2 | 6 | 10 | 8 | 7 | 5 | 5.8 |
| FMRSK | 12 | 10 | 9 | 9 | 6 | 8 | 10 | 9.1 |
| Fudan_zmic | 9 | 11 | 8 | 11 | 9 | 9 | 8 | 9.3 |
| Hilab | 1 | 6 | 13 | 5 | 11 | 12 | 12 | 8.6 |
| Institut_Pasteur_DBC | 15 | 13 | 11 | 12 | 7 | 10 | 11 | 11.3 |
| Neurophet | 16 | 12 | 14 | 14 | 15 | 14 | 16 | 14.4 |
| NVAUTO | 11 | 1 | 12 | 6 | 5 | 3 | 7 | 6.4 |
| Sano | 13 | 14 | 15 | 13 | 12 | 13 | 9 | 12.7 |
| symsense | 8 | 9 | 10 | 4 | 10 | 4 | 4 | 7.0 |
| Uniandes | 6 | 3 | 16 | 7 | 14 | 11 | 15 | 10.3 |
| Xinlab-scut-iai-ahu | 17 | 17 | 17 | 16 | 13 | 15 | 13 | 15.4 |

Also as with FeTA 2021, the top performing teams demonstrated a plateau in performance. This is likely due to the quality of both the SR algorithms and the quality of the manual segmentations. The cortex remains challenging to segment, which supports the importance of maintaining topology in automatic segmentations. The deep GM and brainstem also were more challenging labels to segment. The deep GM is challenging to segment as the structures are not as clearly defined as other structures as there is not a strong demarcation and the difference in intensity from surrounding structures is reduced (unlike in structures such as the ventricles).

In the analysis of topology as an evaluation metric, we demonstrated the importance of considering topology in the assessment and comparison of automatic segmentation methods. In the algorithms ranking, the inclusion of a topology-based metric did not drastically change the final results, although minor updates are observed, and the across-tissue reliability of FMRSK in topological accuracy was rewarded.

Automated fetal brain segmentation has improved since the first FeTA Challenge [11], but there is still room for improvement in the segmentation of certain structures (i.e. cortical grey matter). With the inclusion of just one additional institution (Vienna) into the training dataset, algorithms can improve their generalizability. Future research directions should focus on enhancing the generalizability of the methods, including the emerging low-field fetal MRI acquisitions [39], [40], or exploring federated learning approaches. In that context, conducting a more comprehensive evaluation of the impact of data augmentation and possible biases due to super-resolution reconstruction methods would be very valuable. Furthermore, addressing challenges associated with inaccurate voxel-wise annotations and establishing standards of minimal image quality requirements [41]–[43] should be a priority.

## ACKNOWLEDGMENT

*indicates co-last author. K. Payette is with the Center for MR Research, University Children's Hospital Zurich, University of Zurich, Zurich, Switzerland and the Department of Perinatal Imaging & Health, King's College London, London, UK (email: kelly.payette@kispi.uzh.ch). C. Steger, A. Jakab, H. Ji, L. Wilhelmi, A. G. Gennari, and A. Dändliker are with the Center for MR Research, University Children's Hospital Zurich, University of Zurich, Zurich, Switzerland. C. Steger, A. Jakab, and H. Ji are also with the Neuroscience Center Zurich, University of Zurich, Zurich, Switzerland. A. G. Gennari is also with the Department of Neuropediatrics, University Children's Hospital Zurich, University of Zurich, Zurich, Switzerland. (emails: celine.steger@kispi.uzh.ch, andras.jakab@kispi.uzh.ch, hiu.ji@kispi.uzh.ch, Luca.Wilhelmi@gmail.com, AntonioGiulio.Gennari@kispi.uzh.ch,Aline.Daendliker@kispi.uzh.ch). R. Licandro is with the Laboratory for Computational Neuroimaging, Athinoula A. Martinos Center for Biomedical Imaging, Massachusetts General Hospital/Harvard Medical School, Boston, MA, USA and the Computational Imaging Research Lab , Department of Biomedical Imaging and Image-guided Therapy, , Medical University of Vienna, Vienna, Austria (email: roxane.licandro@meduniwien.ac.at). P. de Dumast and M. Bach Cuadra are with the Medical Image Analysis Laboratory, Department of Diagnostic and Interventional Radiology, Lausanne University Hospital and University of Lausanne, Lausanne, Switzerland and the CIBM Center for Biomedical Imaging, Lausanne, Switzerland (emails: priscille.guerrier-de-dumast@cpe.fr, meritxell.bachcuadra@unil.ch). H.B. Li is with Athinoula A. Martinos Center for Biomedical Imaging, Harvard Medical School (email: holi2@mgh.harvard.edu). M. Barkovich is with the Department of Radiology and Biomedical Imaging, University of California, San Francisco, UCSF Benioff Children's Hospital, San Francisco, USA (email: Matthew.Barkovich@ucsf.edu). L. Liu, C. Chen, and C. Ouyang are with Imperial College London, London, UK. C. Chen is also with the Department of Engineering Science, University of Oxford and the Department of Computer Science, University of Sheffield (emails: liu.li20@imperial.ac.uk, chen.chen15@imperial.ac.uk, c.ouyang@imperial.ac.uk). M. Dannecker is with the Technical University of Munich, Munich, Germany (email: m.dannecker@tum.de). N. McConnell was with Brunel University London, and is now with University College London, UK (email: niccolo.mcconnell.17@ucl.ac.uk). A. Miron, and Y. Li are with Brunel University London, London, UK (emails: alina.miron@brunel.ac.uk, yongmin.li@brunel.ac.uk). A. Uus, I. Grigorescu, and P. Ramierz Gilliland are with the Biomedical Engineering Department, School of Biomedical Engineering & Imaging Sciences, King's College London, London, UK (emails: alena.uus@kcl.ac.uk, irina.grigorescu@kcl.ac.uk, paula.ramirez_gilliland@kcl.ac.uk). Md. M. R. Siddiquee, D. Xu, and A. Myronenko are with NVIDIA. Md. M. R. Siddiquee is also with Arizona State University, USA (emails: mrahmans@asu.edu, daguangx@nvidia.com, amyronenko@nvidia.com). H. Wang and Z. Huang are with Shanghai Jiao Tong University, Shanghai, China (emails: small_dark@sjtu.edu.cn, ziyanhuang@sjtu.edu.cn). J. Ye is with the Shanghai AI Lab, Shanghai, China (email: yejin16@mails.ucas.ac.cn). M. Alenyà, V. Comte, and O. Camara are with BCN-MedTech, Department of Information and Communications Technologies, Universitat Pompeu Fabra, Barcelona, Spain (emails: mireia.alenya@upf.edu, oscar.camara@upf.edu, valentin.comte@upf.edu). J-B Masson and C. Godard are with the Institute Pasteur, Paris, France (emails: jbmasson@pasteur.fr, charlotte.godard@pasteur.fr). A. Nilsson is with the École Polytechnique, Palaiseau, France (email: astrid.nilsson@polytechnique.edu). M. Mazher is with the Centre for Medical Image Computing, Department of Computer Science, University College London, UK (email: moona.mazher@gmail.com). A. Qayyum is with the National Heart and Lung Institute, Imperial College London, London, UK (email: engr.qayyum@gmail.com). Y. Gao, H. Zhou, and S. Gao are with the School of Data Science, Fudan University, Fudan, China (emails: ybgao22@m.fudan.edu.cn, 17307110414@fudan.edu.cn, shqgao@163.com). J. Fu, G.

Dong, and G. Wang are with the School of Mechanical and Electrical Engineering, University of Electronic Science and Technology of China, Chengdu, China (emails: fujia98914@gmail.com, dgm0012@163.com, guotai.wang@uestc.edu.cn). Z. Rieu, H. Yang, and M. Lee are with NEUROPHET Research Institute, Republic of Korea (emails: clarence@neurophet.com, hsyang@neurophet.com, minwoo@neurophet.com). S. Płotka and M. K. Grzeszczyk are with the Sano Centre for Computational Medicine, Cracow, Poland. S. Płotka is also with the Quantitative Healthcare Analysis (qurAI) group, Informatics Institute, University of Amsterdam, Amsterdam, The Netherlands and the Department of Biomedical Engineering and Physics, Amsterdam University Medical Center, Amsterdam, The Netherlands (emails: s.plotka@saonscience.org, m.grzeszczyk@sanoscience.org). A. Sitek is with the Center for Advanced Medical Computing and Simulation, Massachusetts General Hospital, Harvard Medical School, Boston, MA, USA (email: asitek@mgh.harvard.edu). L. Vargas Daza, S. Usma, and P. Arbelaez are with the Center for Research and Formation in Artificial Intelligence, Universidad de los Andes, Colombia (emails:lf.vargas10@uniandes.edu.co, as.usma@uniandes.edu.co, pa.arbelaez@uniandes.edu.co). W. Lu, W. Zhang, and J. Liang are with the School of Electronic and Information Engineering, South China University of Technology, China. Additionally, W. Lu is with AHU-IAI AI Joint Laboratory and AnHui University, China (emails: 1148434471@qq.com, eewenhaozhang@mail.scut.edu.cn,eliangjing@mail.scut.edu.cn). R. Valabregue is with CENIR, ICM, INSERM U 1127, CNRS UMR F7225, Sorbonne Université. Paris France (email: romain.valabregue@upmc.fr). A. A. Joshi, K. N. Nayak, and R. M. Leahy are with Signal and Image Processing Institute, University of Southern California, Los Angeles, CA USA (emails: ajoshi@usc.edu, knayak@usc.edu, leahy@sipi.usc.edu). A. Jakovčić, M. Klaić, A. Adžić, P. Marković, G. Grabarić, and M. Rados are with the Croatian Institute for Brain Research, School of Medicine, University of Zagreb, Zagreb, Croatia(emails: anton.jakovcic97@gmail.com, melita.klaic@gmail.com, anaa.adzic@gmail.com, pavelm990@gmail.com, grabaricgracia@gmail.com, mrados3@yahoo.com). L. Vasung is with the Division of Newborn Medicine, Department of Pediatrics, Boston Children's Hospital and the.Department of Pediatrics, Harvard Medical School Boston, USA (email: Lana.Vasung@gmail.com). G. Kasprian is with the Department of Biomedical Imaging and Image-guided Therapy, Division of Neuroradiology and Musculoskeletal Radiology Medical University of Vienna (email: gregor.kasprian@meduniwien.ac.at). G. Dovjak is with the Department of Biomedical Imaging and Image-guided Therapy, Division of General and Paediatric Radiology, Medical University of Vienna (email: gregor.dovjak@gmx.net)

## REFERENCES

[1] A. Gholipour, J. A. Estroff, C. E. Barnewolt, R. L. Robertson, P. E. Grant, B. Gagoski, S. K. Warfield, O. Afacan, S. A. Connolly, J. J. Neil, A. Wolfberg, and R. V. Mulkern, "Fetal MRI: A Technical Update with Educational Aspirations," *Concepts Magn Reson Part A Bridg Educ Res*, vol. 43, no. 6, pp. 237–266, Nov. 2014.

[2] W. Yan, L. Huang, L. Xia, S. Gu, F. Yan, Y. Wang, and Q. Tao, "MRI Manufacturer Shift and Adaptation: Increasing the Generalizability of Deep Learning Segmentation for MR Images Acquired with Different Scanners," *Radiology: Artificial Intelligence*, vol. 2, no. 4, p. e190195, Jul. 2020.

[3] B. Glocker, R. Robinson, D. C. Castro, Q. Dou, and E. Konukoglu, "Machine Learning with Multi-Site Imaging Data: An Empirical Study on the Impact of Scanner Effects," *arXiv:1910.04597 [cs, eess, q-bio]*, Oct. 2019.

[4] J. R. Zech, M. A. Badgeley, M. Liu, A. B. Costa, J. J. Titano, and E. K. Oermann, "Variable generalization performance of a deep learning model to detect pneumonia in chest radiographs: A cross-sectional study," *PLOS Medicine*, vol. 15, no. 11, p. e1002683, Nov. 2018.

[5] H. Guan and M. Liu, "Domain Adaptation for Medical Image Analysis: A Survey," *IEEE Transactions on Biomedical Engineering*, vol. 69, no. 3, pp. 1173–1185, Mar. 2022.

[6] V. M. Campello, P. Gkontra, C. Izquierdo, C. Martín-Isla, A. Sojoudi, P. M. Full, K. Maier-Hein, Y. Zhang, Z. He, J. Ma, M. Parreño, A. Albiol, F. Kong, S. C. Shadden, J. C. Acero, V. Sundaresan, M. Saber, M. Elattar, H. Li, B. Menze, F. Khader, C. Haarburger, C. M. Scannell, M. Veta, A. Carscadden, K. Punithakumar, X. Liu, S. A. Tsaftaris, X. Huang, X. Yang, L. Li, X. Zhuang, D. Viladés, M. L. Descalzo, A. Guala, L. L. Mura, M. G. Friedrich, R. Garg, J. Lebel, F. Henriques, M. Karakas, E. Çavuş, S. E. Petersen, S. Escalera, S. Seguí, J. F. Rodríguez-Palomares, and K. Lekadir, "Multi-Centre, Multi-Vendor and Multi-Disease Cardiac Segmentation: The M amp;Ms Challenge," *IEEE Transactions on Medical Imaging*, vol. 40, no. 12, pp. 3543–3554, Dec. 2021.

[7] Y. Sun, K. Gao, Z. Wu, G. Li, X. Zong, Z. Lei, Y. Wei, J. Ma, X. Yang, X. Feng, L. Zhao, T. Le Phan, J. Shin, T. Zhong, Y. Zhang, L. Yu, C. Li, R. Basnet, M. O. Ahmad, M. N. S. Swamy, W. Ma, Q. Dou, T. D. Bui, C. B. Noguera, B. Landman, I. H. Gotlib, K. L. Humphreys, S. Shultz, L. Li, S. Niu, W. Lin, V. Jewells, D. Shen, G. Li, and L. Wang, "Multi-Site Infant Brain Segmentation Algorithms: The iSeg-2019 Challenge," *IEEE Transactions on Medical Imaging*, vol. 40, no. 5, pp. 1363–1376, May 2021.

[8] M. Bento, I. Fantini, J. Park, L. Rittner, and R. Frayne, "Deep Learning in Large and Multi-Site Structural Brain MR Imaging Datasets," *Frontiers in Neuroinformatics*, vol. 15, 2022.

[9] T. Eche, L. H. Schwartz, F.-Z. Mokrane, and L. Dercle, "Toward Generalizability in the Deployment of Artificial Intelligence in Radiology: Role of Computation Stress Testing to Overcome Underspecification," *Radiol Artif Intell*, vol. 3, no. 6, p. e210097, Oct. 2021.

[10] C. Martín-Isla, V. M. Campello, C. Izquierdo, K. Kushibar, C. Sendra-Balcells, P. Gkontra, A. Sojoudi, M. J. Fulton, T. W. Arega, K. Punithakumar, L. Li, X. Sun, Y. Al Khalil, D. Liu, S. Jabbar, S. Queirós, F. Galati, M. Mazher, Z. Gao, M. Beetz, L. Tautz, C. Galazis, M. Varela, M. Hüllebrand, V. Grau, X. Zhuang, D. Puig, M. A. Zuluaga, H. Mohy-ud-Din, D. Metaxas, M. Breeuwer, R. J. van der Geest, M. Noga, S. Bricq, M. E. Rentschler, A. Guala, S. E. Petersen, S. Escalera, J. F. R. Palomares, and K. Lekadir, "Deep Learning Segmentation of the Right Ventricle in Cardiac MRI: The M&Ms Challenge," *IEEE Journal of Biomedical and Health Informatics*, vol. 27, no. 7, pp. 3302–3313, Jul. 2023.

[11] K. Payette, H. B. Li, P. de Dumast, R. Licandro, H. Ji, M. M. R. Siddiquee, D. Xu, A. Myronenko, H. Liu, Y. Pei, L. Wang, Y. Peng, J. Xie, H. Zhang, G. Dong, H. Fu, G. Wang, Z. Rieu, D. Kim, H. G. Kim, D. Karimi, A. Gholipour, H. R. Torres, B. Oliveira, J. L. Vilaça, Y. Lin, N. Avisdris, O. Ben-Zvi, D. B. Bashat, L. Fidon, M. Aertsen, T. Vercauteren, D. Sobotka, G. Langs, M. Alenyà, M. I. Villanueva, O. Camara, B. S. Fadida, L. Joskowicz, L. Weibin, L. Yi, L. Xuesong, M. Mazher, A. Qayyum, D. Puig, H. Kebiri, Z. Zhang, X. Xu, D. Wu, K. Liao, Y. Wu, J. Chen, Y. Xu, L. Zhao, L. Vasung, B. Menze, M. B. Cuadra, and A. Jakab, "Fetal brain tissue annotation and segmentation challenge results," *Medical Image Analysis*, vol. 88, p. 102833, Aug. 2023.

[12] L. Maier-Hein, A. Reinke, M. Kozubek, A. L. Martel, T. Arbel, M. Eisenmann, H. Hanbury, P. Jannin, H. Müller, S. Onogur, J. Saez-Rodriguez, B. van Ginneken, A. Kopp-Schneider, and B. A. Landman, "BIAS: Transparent reporting of biomedical image analysis challenges," *Medical Image Analysis*, vol. 66, p. 101796, Dec. 2020.

[13] K. Payette, P. de Dumast, H. Kebiri, I. Ezhov, J. C. Paetzold, S. Shit, A. Iqbal, R. Khan, R. Kottke, P. Grehten, H. Ji, L. Lanczi, M. Nagy, M. Beresova, T. D. Nguyen, G. Natalucci, T. Karayannis, B. Menze, M. Bach Cuadra, and A. Jakab, "An automatic multi-tissue human fetal brain segmentation benchmark using the Fetal Tissue Annotation Dataset," *Sci Data*, vol. 8, no. 1, p. 167, Jul. 2021.

[14] A. Reinke, M. Eisenmann, S. Onogur, M. Stankovic, P. Scholz, P. M. Full, H. Bogunovic, B. A. Landman, O. Maier, B. Menze, G. C. Sharp, K. Sirinukunwattana, S. Speidel, F. van der Sommen, G. Zheng, H. Müller, M. Kozubek, T. Arbel, A. P. Bradley, P. Jannin, A. Kopp-Schneider, and L. Maier-Hein, "How to Exploit Weaknesses in Biomedical Challenge Design and Organization," in *Medical Image Computing and Computer Assisted Intervention – MICCAI 2018*, 2018, pp. 388–395.

[15] H. Lajous, C. W. Roy, T. Hilbert, P. de Dumast, S. Tourbier, Y. Alemán-Gómez, J. Yerly, T. Yu, H. Kebiri, K. Payette, J.-B. Ledoux, R. Meuli, P. Hagmann, A. Jakab, V. Dunet, M. Koob, T. Kober, M. Stuber, and M. Bach Cuadra, "A Fetal Brain magnetic resonance Acquisition Numerical phantom (FaBiAN)," *Sci Rep*, vol. 12, no. 1, p. 8682, May 2022.

[16] K. Payette, C. Steger, R. Licandro, P. de Dumast, H. B. Li, M. Barkovich, L. Li, M. Dannecker, C. Chen, C. Ouyang, N. McConnell, A. Miron, Y. Li, A. Uus, I. Grigorescu, P. Ramirez Gilliland, M. M. Rahman Siddiquee, D. Xu, A. Myronenko, H. Wang, Z. Huang, J. Ye, M. Alenyà, V. Comte, O. Camara, J.-B. Masson, A. Nilsson, C. Godard, M. Mazher, A. Qayyum, Y. Gao, H. Zhou, S. Gao, J. Fu, G. Dong, G. Wang, Z. Rieu, H. Yang, M. Lee, S. Płotka, M. K. Grzeszczyk, A. Sitek, L. Vargas Daza, S. Usma, P. Arbelaez, W. Lu, W. Zhang, J. Liang, R. Valabregue, A. A. Joshi, K. N. Nayak, R. M. Leahy, L. Wilhelmi, A. Dändliker, H. Ji, A. G. Gennari, A. Jakovčić, M. Klaić, A. Adžić, P. Marković, G. Grabarić, G. Kasprian, G. Dovjak, M. Rados, L. Vasung, M. Bach Cuadra, and A. Jakab, "Supplementary Information for the Fetal Tissue Annotation 2022 Challenge Results," Feb. 2024.

[17] K. Payette, C. Steger, P. de Dumast, A. Jakab, M. B. Cuadra, L. Vasung, R. Licandro, M. Barkovich, and H. Li, "Fetal Tissue Annotation Challenge," Mar. 2022.

[18] K. Payette, H. Li, P. de Dumast, R. Licandro, H. Ji, M. M. R. Siddiquee, D. Xu, A. Myronenko, H. Liu, Y. Pei, L. Wang, Y. Peng, J. Xie, H. Zhang, G. Dong, H. Fu, G. Wang, Z. Rieu, D. Kim, H. G. Kim, D. Karimi, A. Gholipour, H. R. Torres, B. Oliveira, J. L. Vilaça, Y. Lin, N. Avisdris, O. Ben-Zvi, D. B. Bashat, L. Fidon, M. Aertsen, T. Vercauteren, D. Sobotka, G. Langs, M. Alenyà, M. I. Villanueva, O. Camara, B. S. Fadida, L. Joskowicz, L. Weibin, L. Yi, L. Xuesong, M. Mazher, A. Qayyum, D. Puig, H. Kebiri, Z. Zhang, X. Xu, D. Wu, K. Liao, Y. Wu, J. Chen, Y. Xu, L. Zhao, L. Vasung, B. Menze, M. B. Cuadra, and A. Jakab, "Fetal Brain Tissue Annotation and Segmentation Challenge Results." arXiv, 20-Apr-2022.

[19] E. Schwartz, M. C. Diogo, S. Glatter, R. Seidl, P. C. Brugger, G. M. Gruber, H. Kiss, K.-H. Nenning, IRC5 consortium, G. Langs, D. Prayer, and G. Kasprian, "The Prenatal Morphomechanic Impact of Agenesis of the Corpus Callosum on Human Brain Structure and Asymmetry," *Cerebral Cortex*, vol. 31, no. 9, pp. 4024–4037, Sep. 2021.

[20] P. Coupe, P. Yger, S. Prima, P. Hellier, C. Kervrann, and C. Barillot, "An Optimized Blockwise Nonlocal Means Denoising Filter for 3-D Magnetic Resonance Images," *IEEE Transactions on Medical Imaging*, vol. 27, no. 4, pp. 425–441, Apr. 2008.

[21] Chao Dong, Chen Change Loy, and Xiaoou Tang, "Accelerating the Super-Resolution Convolutional Neural Network," in *In Computer Vision - ECCV 2016, 9906 LNCS*, 2016, pp. 391–407.

[22] M. Ebner, G. Wang, W. Li, M. Aertsen, P. A. Patel, R. Aughwane, A. Melbourne, T. Doel, S. Dymarkowski, P. De Coppi, A. L. David, J. Deprest, S. Ourselin, and T. Vercauteren, "An automated framework for localization, segmentation and super-resolution reconstruction of fetal brain MRI," *NeuroImage*, vol. 206, p. 116324, Feb. 2020.

[23] A. Gholipour, C. Rollins, C. Velasco-Annis, A. Ouaalam, A. Akhondi-Asl, O. Afacan, C. Ortinau, S. Clancy, C. Limperopoulos, E. Yang, J. Estroff, and S. Warfield, "A normative spatiotemporal MRI atlas of the fetal brain for automatic segmentation and analysis of early brain growth," *Scientific Reports*, vol. 7, Jan. 2017.

[24] S. Tourbier, X. Bresson, P. Hagmann, J.-P. Thiran, R. Meuli, and M. B. Cuadra, "An efficient total variation algorithm for super-resolution in fetal brain MRI with adaptive regularization," *Neuroimage*, vol. 118, pp. 584–597, Sep. 2015.

[25] A. A. Taha and A. Hanbury, "Metrics for evaluating 3D medical image segmentation: analysis, selection, and tool," *BMC Med Imaging*, vol. 15, Aug. 2015.

[26] M. Wiesenfarth, A. Reinke, B. A. Landman, M. Eisenmann, L. A. Saiz, M. J. Cardoso, L. Maier-Hein, and A. Kopp-Schneider, "Methods and open-source toolkit for analyzing and visualizing challenge results," *Sci Rep*, vol. 11, p. 2369, Jan. 2021.

[27] G. Rote and G. Vegter, "Computational Topology: An Introduction," in *Effective Computational Geometry for Curves and Surfaces*, J.-D. Boissonnat and M. Teillaud, Eds. Berlin, Heidelberg: Springer, 2006, pp. 277–312.

[28] H. Dou, D. Karimi, C. K. Rollins, C. M. Ortinau, L. Vasung, C. Velasco-Annis, A. Ouaalam, X. Yang, D. Ni, and A. Gholipour, "A Deep Attentive Convolutional Neural Network for Automatic Cortical Plate Segmentation in Fetal MRI," *IEEE Transactions on Medical Imaging*, vol. 40, no. 4, pp. 1123–1133, Apr. 2021.

[29] A. E. Fetit, A. Alansary, L. Cordero-Grande, J. Cupitt, A. B. Davidson, A. D. Edwards, J. V. Hajnal, E. Hughes, K. Kamnitsas, V. Kyriakopoulou, A. Makropoulos, P. A. Patkee, A. N. Price, M. A. Rutherford, and D. Rueckert, "A deep learning approach to segmentation of the developing cortex in fetal brain MRI with minimal manual labeling," in *Proceedings of the Third Conference on Medical Imaging with Deep Learning*, 2020, vol. 121, pp. 241–261.

[30] J. Hong, H. J. Yun, G. Park, S. Kim, C. T. Laurentys, L. C. Siqueira, T. Tarui, C. K. Rollins, C. M. Ortinau, P. E. Grant, J.-M. Lee, and K. Im, "Fetal Cortical Plate Segmentation Using Fully Convolutional Networks With Multiple Plane Aggregation," *Frontiers in Neuroscience*, vol. 14, p. 1226, 2020.

[31] B. Caldairou, N. Passat, P. Habas, C. Studholme, M. Koob, J.-L. Dietemann, and F. Rousseau, "Segmentation of the cortex in fetal MRI using a topological model," in *2011 IEEE International Symposium on Biomedical Imaging: From Nano to Macro*, 2011, pp. 2045–2048.

[32] P. de Dumast, H. Kebiri, C. Atat, V. Dunet, M. Koob, and M. B. Cuadra, "Segmentation of the Cortical Plate in Fetal Brain MRI with a Topological Loss," in *Uncertainty for Safe Utilization of Machine Learning in Medical Imaging, and Perinatal Imaging, Placental and Preterm Image Analysis: 3rd International Workshop, UNSURE 2021, and 6th International Workshop, PIPPI 2021, Held in Conjunction with MICCAI 2021, Strasbourg, France, October 1, 2021, Proceedings*, Berlin, Heidelberg, 2021, pp. 200–209.

[33] F. Isensee, P. F. Jaeger, S. A. A. Kohl, J. Petersen, and K. H. Maier-Hein, "nnU-Net: a self-configuring method for deep learning-based biomedical image segmentation," *Nat Methods*, vol. 18, no. 2, pp. 203–211, Feb. 2021.

[34] MONAI Consortium, "MONAI: Medical Open Network for AI." Mar-2020.

[35] M. Kuklisova-Murgasova, G. Quaghebeur, M. A. Rutherford, J. V. Hajnal, and J. A. Schnabel, "Reconstruction of fetal brain MRI with intensity matching and complete outlier removal," *Med Image Anal*, vol. 16, no. 8, pp. 1550–1564, Dec. 2012.

[36] C. Chen, C. Qin, H. Qiu, C. Ouyang, S. Wang, L. Chen, G. Tarroni, W. Bai, and D. Rueckert, "Realistic Adversarial Data Augmentation for MR Image Segmentation," in *Medical Image Computing and Computer Assisted Intervention – MICCAI 2020*, Cham, 2020, pp. 667–677.

[37] C. Chen, Z. Li, C. Ouyang, M. Sinclair, W. Bai, and D. Rueckert, "MaxStyle: Adversarial Style Composition for Robust Medical Image Segmentation," in *Medical Image Computing and Computer Assisted Intervention – MICCAI 2022: 25th International Conference, Singapore, September 18–22, 2022, Proceedings, Part V*, Berlin, Heidelberg, 2022, pp. 151–161.

[38] P. Chattopadhyay, Y. Balaji, and J. Hoffman, "Learning to Balance Specificity and Invariance for In and Out of Domain Generalization," in *16th European Conference, Glasgow, UK, August 23–28, 2020, Proceedings, Part IX 16. Springer International Publishing, 2020*, Glasgow, UK, 2020.

[39] J. Aviles Verdera, L. Story, M. Hall, T. Finck, A. Egloff, P. T. Seed, S. J. Malik, M. A. Rutherford, J. V. Hajnal, R. Tomi-Tricot, and J. Hutter, "Reliability and Feasibility of Low-Field-Strength Fetal MRI at 0.55 T during Pregnancy," *Radiology*, vol. 309, no. 1, p. e223050, Oct. 2023.

[40] K. Payette, A. Uus, J. Aviles Verdera, C. Avena Zampieri, M. Hall, L. Story, M. Deprez, M. A. Rutherford, J. V. Hajnal, S. Ourselin, R. Tomi-Tricot, and J. Hutter, "An Automated Pipeline for Quantitative T2* Fetal Body MRI and Segmentation at Low Field," in *Medical Image*

*Computing and Computer Assisted Intervention – MICCAI 2023*, Cham, 2023, pp. 358–367.

[41] T. Sanchez, O. Esteban, Y. Gomez, E. Eixarch, and M. B. Cuadra, "FetMRQC: Automated Quality Control for Fetal Brain MRI," in *Perinatal, Preterm and Paediatric Image Analysis: 8th International Workshop, PIPPI 2023, Held in Conjunction with MICCAI 2023, Vancouver, BC, Canada, October 12, 2023, Proceedings*, Berlin, Heidelberg, 2023, pp. 3–16.

[42] K. Payette, R. Kottke, and A. Jakab, "Efficient multi-class fetal brain segmentation in high resolution MRI reconstructions with noisy labels," in *Medical Ultrasound, and Preterm, Perinatal and Paediatric Image Analysis*, 2020, pp. 295–304.

[43] D. Karimi, C. K. Rollins, C. Velasco-Annis, A. Ouaalam, and A. Gholipour, "Learning to segment fetal brain tissue from noisy annotations," *Medical Image Analysis*, vol. 85, p. 102731, Apr. 2023.